\def\href#1#2{{#2}}
\documentclass[aps,prb,floats,amssymb,twocolumn,letterpaper,groupedaddress,
showpacs,floatfix,twoside,tightenlines,superscriptaddress]{revtex4}
\usepackage{graphicx}
\usepackage{verbatim}
\usepackage{amssymb,amsmath}
\usepackage{epsfig}
\begin{document}

\title{Step Position Distributions and the Pairwise Einstein Model for 
       Steps on Crystal Surfaces}
\author{Amber N.\ Benson} 
\email[]{amber\_benson@yahoo.com}
\affiliation{Department of Physics and Astronomy, 
     Mississippi State, MS  39762-5167} 
\affiliation{Department of Physics, 
     Texas A \& M University--Commerce, 
     Commerce, TX  75429-3011} 
\author{Howard L.\ Richards} 
\email[Corresponding~author: ]{Howard\_Richards@tamu-commerce.edu}
\homepage[]{http://faculty.tamu-commerce.edu/HRichards/index.html}
\affiliation{Department of Physics, 
     Texas A \& M University--Commerce, 
     Commerce, TX  75429-3011} 
\author{T.~L.\ Einstein}
\email[]{einstein@umd.edu}
\homepage[]{http://www2.physics.umd.edu/~einstein/}
\affiliation{Department of Physics, University of Maryland, College Park, MD 20742-4111}
\date{\today}

\begin{abstract}
The Pairwise Einstein Model (PEM)
of steps not only justifies the use of the Generalized Wigner
Distribution (GWD) for Terrace Width Distributions (TWDs), it also predicts
a specific form for the Step Position Distribution (SPD), {\em i.e.}, the
probability density function for the fluctuations of a step about
its average position. The predicted form of the SPD is well approximated
by a Gaussian with a finite variance. However, the variance of the 
SPD measured from either real surfaces or Monte Carlo simulations 
depends on $\Delta y$, the length of step over which it is 
calculated, with the measured variance diverging in the limit 
$\Delta y \rightarrow \infty$. 
As a result, a length scale $L_{\rm W}$ can be defined as the value of 
$\Delta y$ at which the measured and theoretical SPDs agree.  
Monte Carlo simulations of the terrace-step-kink model 
indicate that $L_{\rm W} \! \approx \! 14.2 \xi_Q$, 
where $\xi_Q$ is the correlation 
length in the direction parallel to the steps,  
independent of the strength of the step-step repulsion.  
$L_{\rm W}$ can also be understood as the length over which a {\em single}
terrace must be sampled for the TWD to bear a ``reasonable'' 
resemblence to the GWD.  
\end{abstract}
\pacs{PACS Number(s):
 05.70.Np, 
 68.55.Jk, 
 68.35.Ct, 
 68.35.-p  
}
\maketitle

\section{Introduction}

A key factor determining the equilibrium morphology of a 
vicinal crystal surface is the interaction between the steps 
on that surface.  In many cases, the elastic and electronic 
contributions to the step-step interaction take the form
\begin{equation}
 \label{eq:InvSquareInteraction}
 V(L) = \frac{A}{L^2} \, ,
\end{equation}
where $A$ determines the strength of the step-step interaction and
$L$ is the distance between steps.  Because this is a typical
step-step interaction, and because it has the remarkable property of
yielding exact solutions to very plausible approximate
theories\cite{Calogero69,Sutherland71,beyond}, we
confine ourselves in this paper to interactions of the form given in
Eq.~(\ref{eq:InvSquareInteraction}).  With this restriction, many of the
quantities discussed in this paper depend only on a single dimensionless
parameter,
\begin{equation}
  \label{eq:deftildeA}
  \tilde{A} \equiv \frac{\tilde{\beta}A}{(k_{\rm B}T)^2} \, ,
\end{equation}
where $\tilde{\beta}$ is the step stiffness,
$k_{\rm B}$ is Boltzmann's constant, and $T$ is the
absolute temperature.

One of the easiest methods\cite{Giesen00,statsteps,SurfSci493_460} 
for experimentally determining the interaction
between steps on a vicinal crystal surface is through the observation of
the Terrace Width Distribution (TWD).
Typically, this has been done by fitting the TWD to a Gaussian, which
is a good approximation and justified by the Gruber-Mullins 
approximation\cite{Gruber67,Bartelt90} 
(analogous to the Einstein model\cite{AshcroftMermin} of solids) 
if the steps strongly repel each other.  
The step-step interaction is
then extracted from the variance of the Gaussian.
Unfortunately, however,
the Gaussian approximation is only good for strongly interacting steps,
and there are conflicting theories\cite{Gruber67,Bartelt90,PM98,IMP98,%
Masson94,Barbier96,LeGoff99} regarding the relationship between
the step-step interaction and the variance.

Over the past decade\cite{Giesen00,statsteps,SurfSci493_460}  
it has become apparent that the so-called
Generalized Wigner Distribution (GWD) provides a much better 
approximation to
the TWD.  The GWD exhibits the positive skew
observed in TWDs from experiments and simulations, and it is a good fit
quantitatively to TWDs produced from Monte Carlo simulations of the
terrace-step-kink (TSK) model.  
More significantly, the GWD can be justified on the basis 
of plausible approximations\cite{beyond,Yancey05}, the most important of 
which is that the interaction and fluctuations of {\em two} adjacent steps 
are explicitly considered; the Gruber-Mullins approximation only explicitly 
considers {\em one} step.  The two steps are kept close to each other by 
a harmonic well, which approximates the interactions with all other 
steps. This model is refered to as the Pairwise Einstein Model (PEM). 
Both the Gruber-Mullins and pairwise Einstein models start by interpreting the 
steps as world-lines of spinless fermions, with the $y$-direcion 
(the average direction of the steps) corresponding to time.  

This paper considers a different statistical measure of the 
vicinal surface:  the Step Position Distribution (SPD). 
In Sec.~\ref{sec-predictions}, we show that the pairwise Einstein model
predicts a Gaussian-like distribution for the {\em position} 
of steps.  In Sec.~\ref{sec-numerical} these predictions are shown to 
compare well with numerical results from simulations of the TSK model, 
at least for systems of the ``right size''.  
The dependence of the SPD on the length of the steps is discussed in 
Sec.~\ref{sec-scaling}; for the purpose of comparison, the dependence 
of the TWD on the length of steps is likewise discussed in 
Sec.~\ref{sec-TWDs}.  Finally, in Sec.~\ref{sec-conclusions} we 
summarize and draw our conclusions. 

\section{Predictions from the Pairwise Einstein Model}
\label{sec-predictions}

As was shown in Ref.~\onlinecite{beyond}, the Generalized Wigner Distribution
can be derived from a phenomenological treatment in which only two
steps are treated explicitly, the rest contributing a ``confinement
potential'' related to the two-dimensional pressure and compressibility
of the system of steps.  We use the usual trick of mapping steps onto
the worldlines of one-dimensional spinless fermions, which in this
case have the Hamiltonian\cite{beyond,Yancey05}
\begin{equation}
  \label{eq:HamX1X2}
    {\cal H} = -\frac{1}{2} \left( \frac{\partial^2}{\partial x_1^2}
                                     + \frac{\partial^2}{\partial x_2^2}
                                \right)
                   + \frac{\tilde{A}}{(x_2-x_1)^2}
                   + \frac{\omega^2}{2} \left( x_1^2 + x_2^2 \right) \, .
\end{equation}
In this dimensionless formulation, we require that
\begin{equation}
  \label{eq:constraint}
  \langle x_2 - x_1 \rangle = 1 \, ;
\end{equation}
this fixes the value of $\omega$ to
\begin{equation}
   \omega = 2 b_\varrho \, ,
\end{equation}
where
\begin{equation}
  \label{eq:brho}
  b_\varrho \equiv \left[\frac{\Gamma\left(\frac{\varrho + 2}{2}\right)}
                              {\Gamma\left(\frac{\varrho + 1}{2}\right)}
                   \right]^2 \, ,
\end{equation}
and
\begin{equation}
  \varrho = 1 + \sqrt{1+ 4\tilde{A}} \, .
\end{equation}
After a change of variables\cite{beyond,Yancey05} to
\begin{eqnarray}
  x_{\rm cm} & = & \frac{x_1 + x_2}{2} \\
           s & = & x_2 - x_1 \, ,
\end{eqnarray}
this Hamiltonian becomes separable\cite{beyond,Yancey05},
\begin{equation}
   \label{eq:HamXcmS}
  {\cal H} = -\left(           \frac{\partial^2}{\partial s^2}
                 + \frac{1}{4} \frac{\partial^2}{\partial x_{\rm cm}^2}
              \right)
            + \frac{\tilde{A}}{s^2}
            + b_\varrho^2 \left( s^2 + 4 x_{\rm cm}^2 \right) \, ,
\end{equation}
and it has the remarkable property that all of the eigenstates are
known.  The only eigenstate of interest to us at present,
however, is the ground state, which can be written\cite{beyond,Yancey05}
\begin{eqnarray}
  \Psi_{0,0}(s,x_{\rm cm}) & = & \left[ a_\varrho^{1/2} s^{\varrho/2}
                              \exp \left(-\frac{b_\varrho s^2}{2}
                                   \right)\right]
         \nonumber \\ & & \times
                                 \left[ \frac{1}{2\sqrt{\pi b_\varrho}}
                                  \exp \left(
                                       -4 b_\varrho x_{\rm cm}^2
                                       \right)\right] \, ,
\end{eqnarray}
where
\begin{equation}
  a_\varrho = \frac{2 b_\varrho^{(\varrho+1)/2}}
                   {\Gamma [(\varrho+1)/2]}
\end{equation}
is a constant of normalization.
The probability of finding the combination a specific combination of relative
separation and ``center of mass'' is, of course, just
$\Psi_{0,0}^{2}(s,x_{\rm cm})$, which
can be rewritten in terms of the original variables $x_1$ and $x_2$:
\begin{eqnarray}
   P(x_1,x_2) & = & \Psi_{0,0}^2(s,x_{\rm cm}) \nonumber \\
              & = & \frac{a_\varrho}{\sqrt{\pi b_\varrho}}
                    (x_2 - x_1)^\varrho
                    \exp[-2b_\varrho (x_1^2 + x_2^2)] \, ,
\end{eqnarray}
subject to the constraint $x_2 \! \geq \! x_1$.
We can integrate out all possible
values of $x_2$ to find the probability density function for $x_1$:
\begin{eqnarray}
  Q_1(x_1) & = & \int_{x_1}^\infty P(x_1,x_2) \, {\rm d}x_2 \nonumber \\
           & = & \frac{a_\varrho}{\sqrt{\pi b_\varrho}}
                 \exp(-2b_\varrho x_1^2) \nonumber \\ & & \mbox{} \times
                 \int_{x_1}^\infty (x_2 - x_1)^\varrho
                 \exp(-2b_\varrho x_2^2) \, {\rm d}x_2 \, .
\end{eqnarray}
As should be expected, the mean value of $x_1$ is $-1/2$ and
the mean value of $x_2$ is $+1/2$, so we define the analytic SPD to be the calculated
probability density function for $x_1 - \langle x_1 \rangle$:
\begin{eqnarray}
   Q(x) & \equiv & Q_1\left(x + \frac{1}{2} \right) \nonumber \\
  \label{eq:Q}
        & = & \frac{a_\varrho}{\sqrt{\pi b_\varrho}}
              \exp\left[-2b_\varrho
                        \left(x + \frac{1}{2}\right)^2 \right]
                 \nonumber \\ & & \mbox{} \times
                 \int_{x}^\infty \left(x_2 - x + \frac{1}{2}\right)^\varrho
              \exp(-2b_\varrho x_2^2) \, {\rm d}x_2 \, .
\end{eqnarray}

Although $Q(x)$ can only be evaluated numerically 
(it can be rewritten as a complicated expression involving hypergeometric functions, 
but this does not seem to be genuinely helpful), it is
straightforward, though tedious, to calculate its moments.
The two most important are the mean, which is
zero by definition, and the variance, which is given by
\begin{eqnarray}
  \label{eq:sigma2}
  \sigma^2_{Q,{\rm W}} & = & \frac{1}{4}\left(\frac{\varrho + 2}
                           {2 b_\varrho} - 1\right) \\
                     & \sim & \frac{3}{8} \varrho^{-1} \, .
\end{eqnarray}
These two moments would be enough to entirely specify the SPD if it
were a Gaussian distribution, which it should be approximately;
the Gruber-Mullins approximation for the TWD, since it concerns the
fluctuations in position of only a single step, can be equally well
interpreted as an approximation for the SPD.  In fact, both the
coefficient of skewness\cite{Roe} and the kurtosis\cite{Roe} of the 
theoretical SPD vanish in
the limit of strong step-step repulsion.
The coefficient of skewness is given asymptotically by
\begin{equation}
   \label{eq:skew}
   \gamma_1 \equiv \frac{\langle (x_1 - \langle x_1 \rangle)^3 \rangle}
                        {\sigma^3_{Q,{\rm W}}}
    \sim -\frac{\sqrt{6}}{18} \varrho^{-1/2} \, ;
\end{equation}
note that the coefficient of skewness would have the
opposite sign if it had
been defined as
$\langle (x_2 - \langle x_2 \rangle)^3 \rangle \sigma_{Q,{\rm W}}^{-3}$.
The kurtosis, which is the same regardless of which step is considered,
is given asymptotically by
\begin{equation}
 \label{eq:kurtosis}
   \gamma_2 \equiv \frac{\langle (x_1 - \langle x_1 \rangle)^4 \rangle}
                        {\sigma^4_{Q,{\rm W}}} - 3
    \sim \frac{1}{12} \varrho^{-2} \, .
\end{equation}

The fact that the kurtosis is not exactly zero is not in itself
surprising; even within the Gruber-Mullins approximation, the
Gaussian distribution is only obtained in the limit of large
$\tilde{A}$.  The symmetry of our original problem
of an infinite number of steps on an infinite vicinal surface,
on the other hand, means that the coefficient of skewness, by contrast,
{\em must} be zero for the original problem.  Any given step on the
surface can be considered ``step~1'', with its downhill neighbor as
``step~2'', or it can be considered ``step~2'', with its
uphill neighbor as ``step~1''; calling it one or the other breaks the
symmetry and permits a nonzero coefficient of skewness.

\section{Comparison with Monte Carlo Simulations}
\label{sec-numerical}

In order to test the applicability of Eq.~(\ref{eq:Q}), we have performed
Monte Carlo simulations of the terrace-step-kink (TSK) model and
measured the SPD for several values of $\tilde{A}$.

The geometry of the simulated systems was as follows.
All simulations were for systems of 30 steps;
the length of each of which was $L_y \! = \! 1000 a$
(where $a$ is the lattice constant) in the average direction of the
steps (the $y$-direction in ``Maryland notation'').
The mean step separation was $\langle L \rangle \! = \! 10a$, and
periodic boundary conditions were applied.

The dynamic used was a local Metropolis update.
The temperature was set at $k_{\rm B}T \! = \! 0.45 \epsilon$,
where $\epsilon$ is the kink energy; in a previous study,
this was approximately the temperature at which TWDs from the
restricted TSK model showed the best agreement with the Generalized
Wigner Distribution.  Each simulation was equilibrated for at least
500~000 Monte Carlo steps per site (MCSS) at the temperature and
value of $\tilde{A}$ at which measurements were taken; the initial
configurations, however, were not typically straight steps, but
steps that had been equilibrated at some other value of $\tilde{A}$.
Data were taken from 1~000
``snapshots,'' taken at intervals of 1~000 MCSS.

Although the terrace width is always an integer multiple of $a$ in the
TSK model, the average step position can be any rational number,
depending only on the size of the simulation.  Since the step position
$x$ is always an integer, the histogram of positions for any given
step need not be symmetric.

\begin{figure}
\epsfxsize=5.7cm
\begin{center}
\centerline{\epsfbox{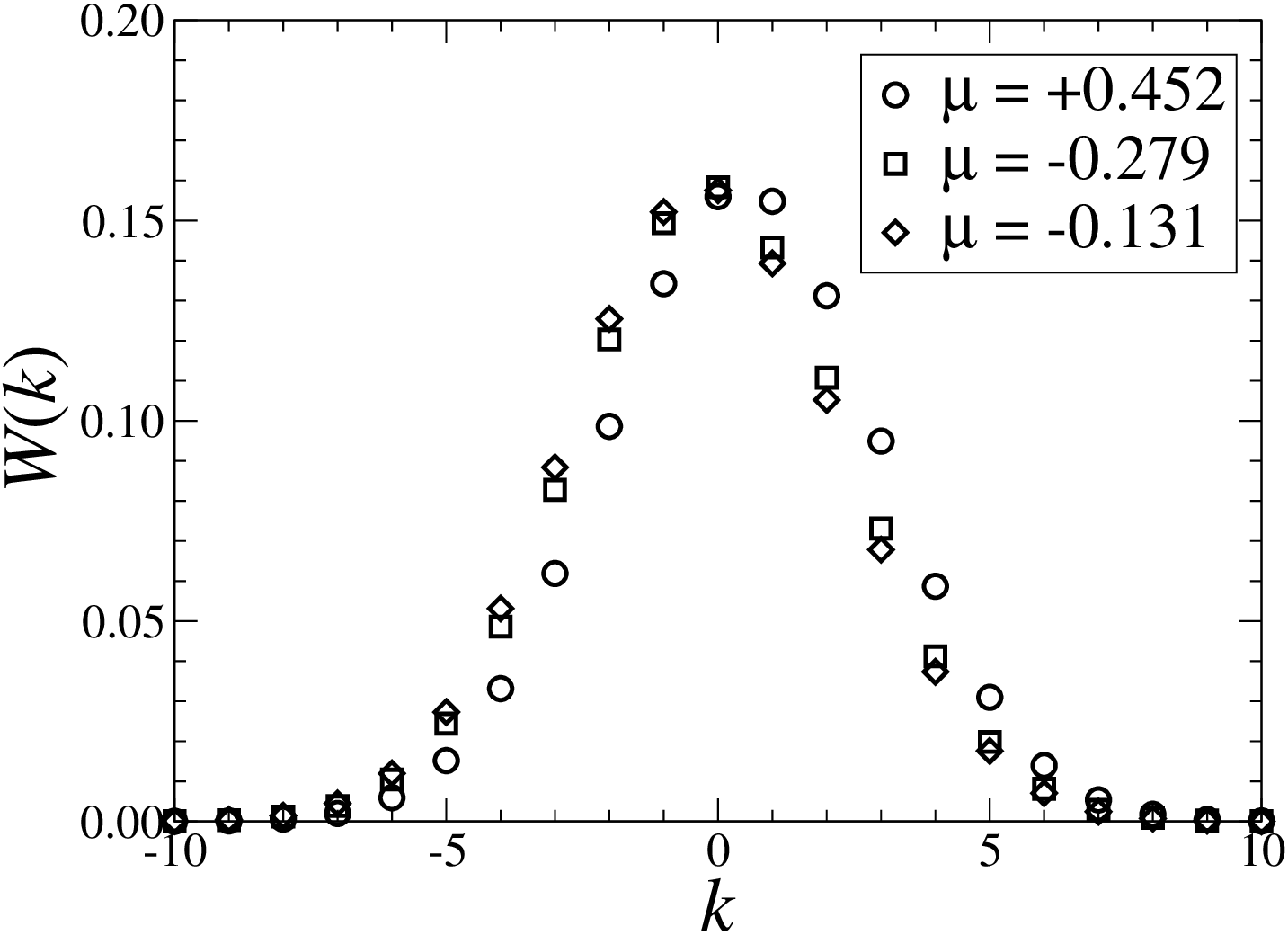}}
\end{center}
\caption[shrt]{
An illustration of problem that can be caused by the variability
of the mean step position when the step position distribution
(SPD) is calculted from numerical or experimental results.
In this example, Gaussian distributions with identical variances
($\sigma^2 \! = \! 2.5^2$) are binned into histograms by means of
Eq.~(\protect\ref{eq:weights}).  The only differences between the
three distributions are the values of $\mu$:
circles, $\mu \! = \! 0.452$;
squares, $\mu \! = \! -0.279$;
diamonds, $\mu \! = \! -0.131$.
}
\protect\label{fig:weights}
\end{figure}

In order to show concretely what this means, consider a situation
in which a Gaussian distribution with mean $\mu$ and variance
$\sigma^2$ is binned into a histogram as follows.  The weight assigned to
each integer $k$ is given by integrating the Gaussian between
$k \! - \! 1/2$ and $k \! + \! 1/2$:
\begin{eqnarray}
   \label{eq:weights}
   W(k) & = & \frac{1}{\sigma \sqrt{2\pi}}
          \int_{k-1/2}^{k+1/2} \exp \left[-\frac{(x-\mu)^2}{2\sigma^2}
                                   \right] \, {\rm d}x \nonumber \\
       & = & \frac{1}{2} \Biggl\{
                {\rm erf}\left[ \frac{k-(1/2)-\mu}{2\sigma}\right]
       \nonumber \\ & & \mbox{}
              - {\rm erf}\left[ \frac{k+(1/2)-\mu}{2\sigma}\right]
                         \Biggr\} \, .
\end{eqnarray}
For our example, we choose $\sigma \! = \! 2.5$ and three ``random''
values of $\mu$ between -0.5 and +0.5.  The results are shown in
Fig.~\ref{fig:weights}.  Clearly none of the histograms is completely
symmetric, and the differences between them are noteworthy.

\begin{figure}
\epsfxsize=5.7cm
\begin{center}
\centerline{\epsfbox{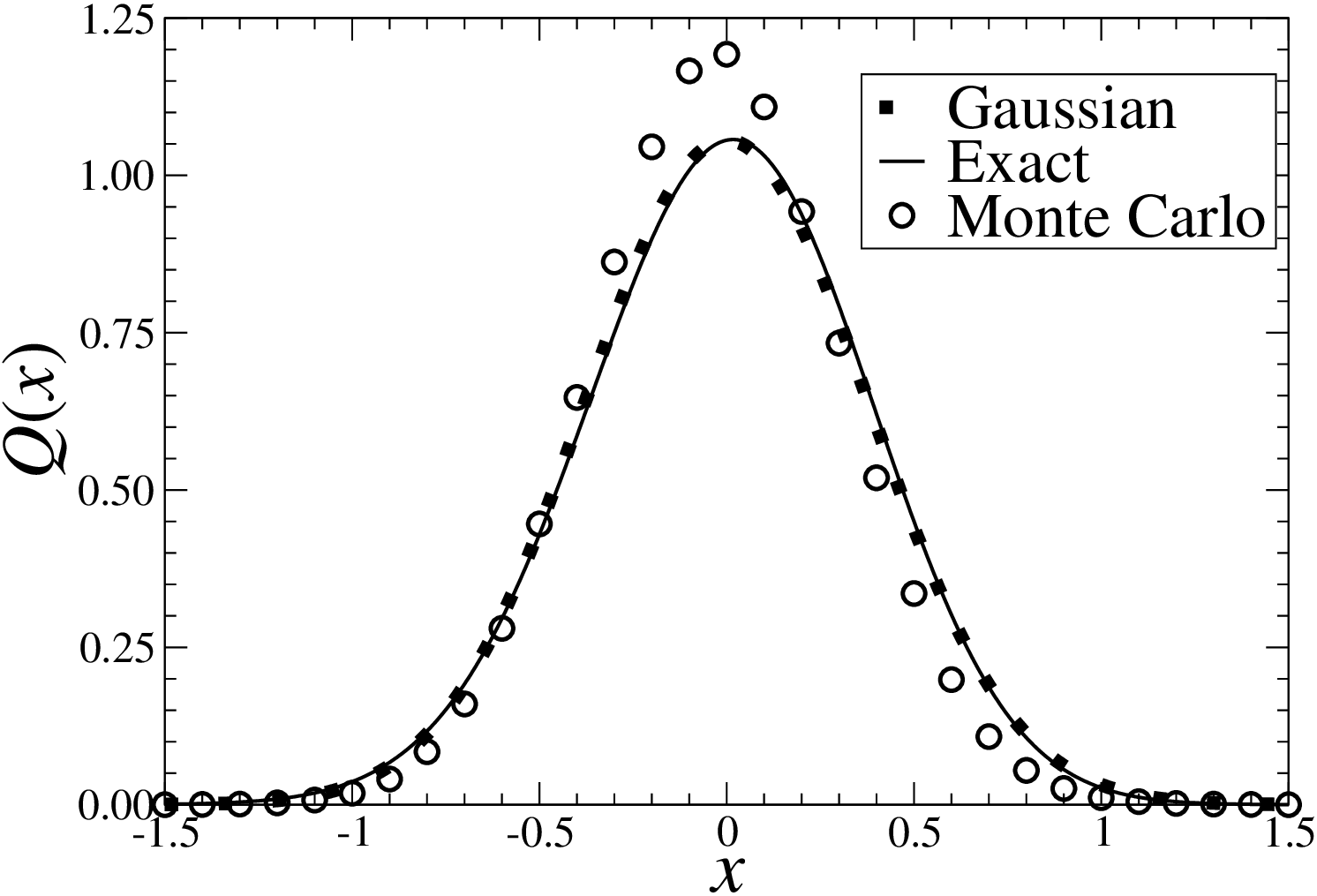}}
\end{center}
\caption[shrt]{
Comparison of the SPD
for $\tilde{A} \! = \! 0$ given by
Eq.~(\protect\ref{eq:Q}) (solid curve) with a histogram SPD from
a Monte Carlo simulation (symbols).
Also shown is a Gaussian (dotted curve) with a mean of
zero and a variance given by Eq.~(\protect\ref{eq:sigma2}).
}
\protect\label{fig:Atilde00}
\end{figure}

\begin{figure}
\epsfxsize=5.7cm
\begin{center}
\centerline{\epsfbox{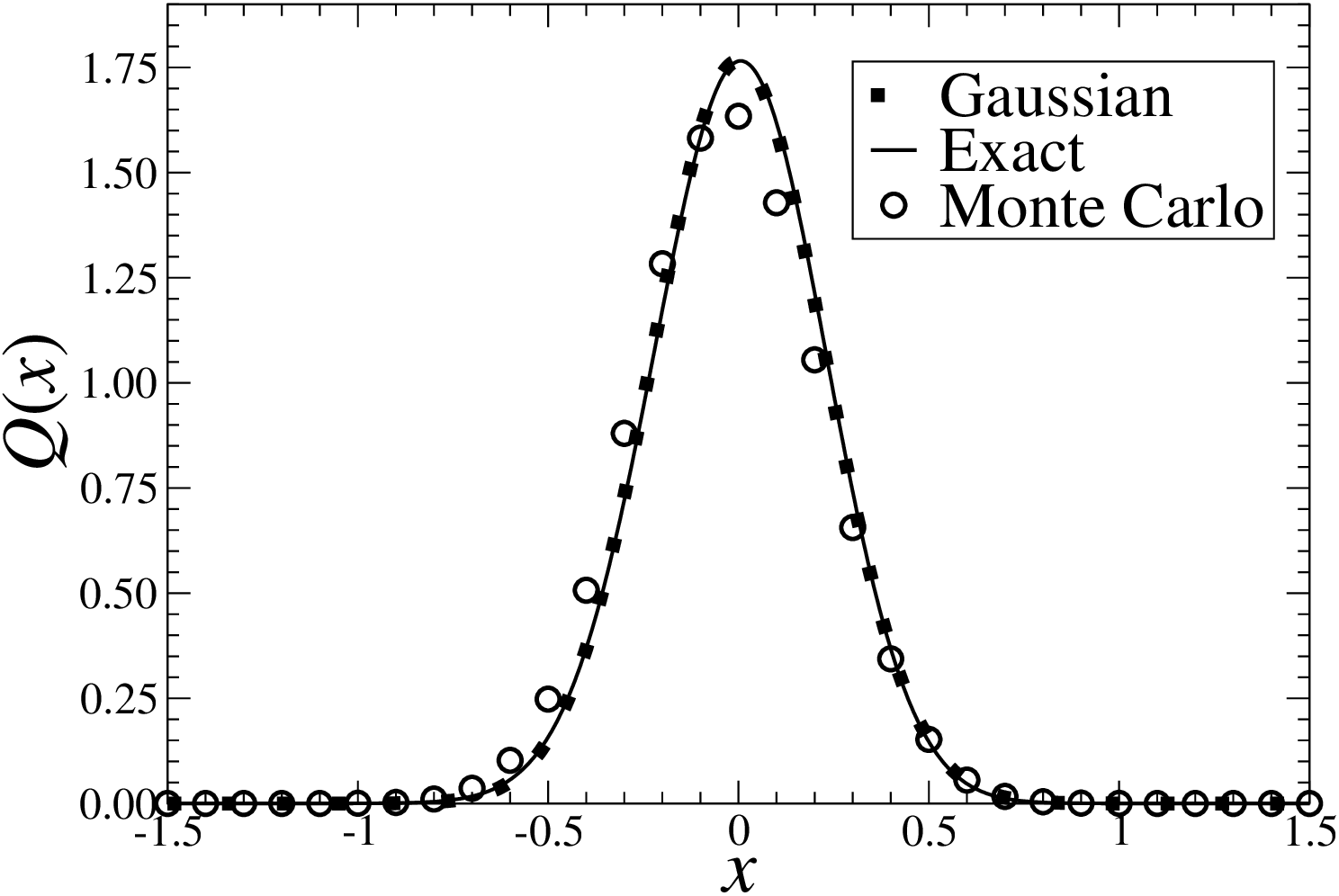}}
\end{center}
\caption[shrt]{
Comparison of the SPD
for $\tilde{A} \! = \! 8$ given by
Eq.~(\protect\ref{eq:Q}) (solid curve) with a histogram SPD from
a Monte Carlo simulation (symbols).
Also shown is a Gaussian (dotted curve) with a mean of
zero and a variance given by Eq.~(\protect\ref{eq:sigma2}).
}
\protect\label{fig:Atilde08}
\end{figure}

Something similar can and does happen when the SPDs are calculated from
Monte Carlo simulations by binning the positions into histograms.
As a result, the statistical uncertainties are considerably larger than
they are for the corresponding TWDs, and the SPDs are not perfectly
symmetric about their peaks, as can be seen in
Figs.~\ref{fig:Atilde00} and \ref{fig:Atilde08}.
Note the qualitative similarities between the Monte Carlo results
(circles) in Figs.~\ref{fig:Atilde00} and \ref{fig:Atilde08}
and the values of $W(k)$ for $\mu \! = \! -0.279$ (squares)
and $\mu \! = \! -0.131$ (diamonds) in Fig.~\ref{fig:weights}.
This agreement suggests that during the process of equlibration, the
majority of the steps moved slightly to the left ({\em i.e.}, uphill).

In spite of this, the agreement of the SPDs calculated from simulations
and the theoretical $Q(x)$ calculated from Eq.~(\ref{eq:Q}) is
reasonably good.  Even more impressive is the agreement between $Q(x)$ and
the Gaussian with zero mean and variance given by Eq.~(\ref{eq:sigma2}).
Although Eqs.~(\ref{eq:skew}) and (\ref{eq:kurtosis}) suggest that the Gaussian
approximation will be increasingly good as $\tilde{A}$ becomes large,
it is clear from the figures that the Gaussian approximation is good
for even for $\tilde{A} \! = \! 0$.

\section{Scaling of the SPD}
\label{sec-scaling}

Although the agreement between Eq.~(\ref{eq:Q}) and the numerical SPDs 
discussed above is highly suggestive, it is clear that actual  
step position distributions must depend on the length $\Delta y$ 
of step over which they are averaged.  This is best demonstrated by 
considering the variance of a measured SPD, which is given by 
\begin{equation}
  \label{eq:varSPD}
   \sigma^2_{Q}(\Delta y) \equiv (\Delta y)^{-1}
	\left\langle
	     \int_{-\Delta y/2}^{\Delta y/2} [x(y)-\overline{x}]^2 dy 
        \right\rangle \, ,  
\end{equation}
where 
\begin{equation}
  \label{eq:avgx}
    \overline{x} \equiv (\Delta y)^{-1}
	\int_{-\Delta y/2}^{\Delta y/2} x(y) dy \, .
\end{equation}
Clearly, $\sigma^2_{Q}(\Delta y)$ is closely related to\cite{role}
\begin{equation}
 \label{eq:gx}
 g_x(\Delta y) \equiv \left\langle 
                    \left[x(\Delta y) - x(0)\right]^2 
                    \right\rangle \, , 
\end{equation}
which characterizes the wandering of an individual 
step\cite{role,Villain81,Villain85,Saam89,Bartelt91}. 

It is tempting to identify $\overline{x}$, the average value of 
$x$ for a particular conformation of a step, with $x(0)$, the 
value of $x$ at the average $y$-position.  This leads to 
\begin{equation}
  \label{eq:approxsig2}
   \sigma^2_{Q}(\Delta y) \approx (\Delta y)^{-1}
	\int_{-\Delta y/2}^{\Delta y/2} g_x(y) dy \, .  
\end{equation}
For small $\Delta y$, 
$g_x \! \approx \! c_1 |\Delta y|$\cite{role,Villain81,Villain85,Saam89,Bartelt91};  
Eq.~(\ref{eq:approxsig2}) implies $\sigma^2_Q \! \approx \! (c_1 / 2) \Delta y$.
Likewise, for large $\Delta y$,\cite{role,Villain81,Villain85,Saam89,Bartelt91}    
\begin{equation}
  \label{eq:asympgx}
  g_x(\Delta y) \approx  c_2 + c_3 ln|\Delta y|
\end{equation}
and Eq.~(\ref{eq:approxsig2}) implies 
\begin{equation}
  \label{eq:logscaledsig2}
   \frac{\sigma^2_Q(\Delta y)}{\sigma^2_{Q,{\rm W}}} \approx c_4 + c_5 ln|\Delta y| \, .
\end{equation}

\begin{figure}
\epsfxsize=5.7cm
\begin{center}
\centerline{\epsfbox{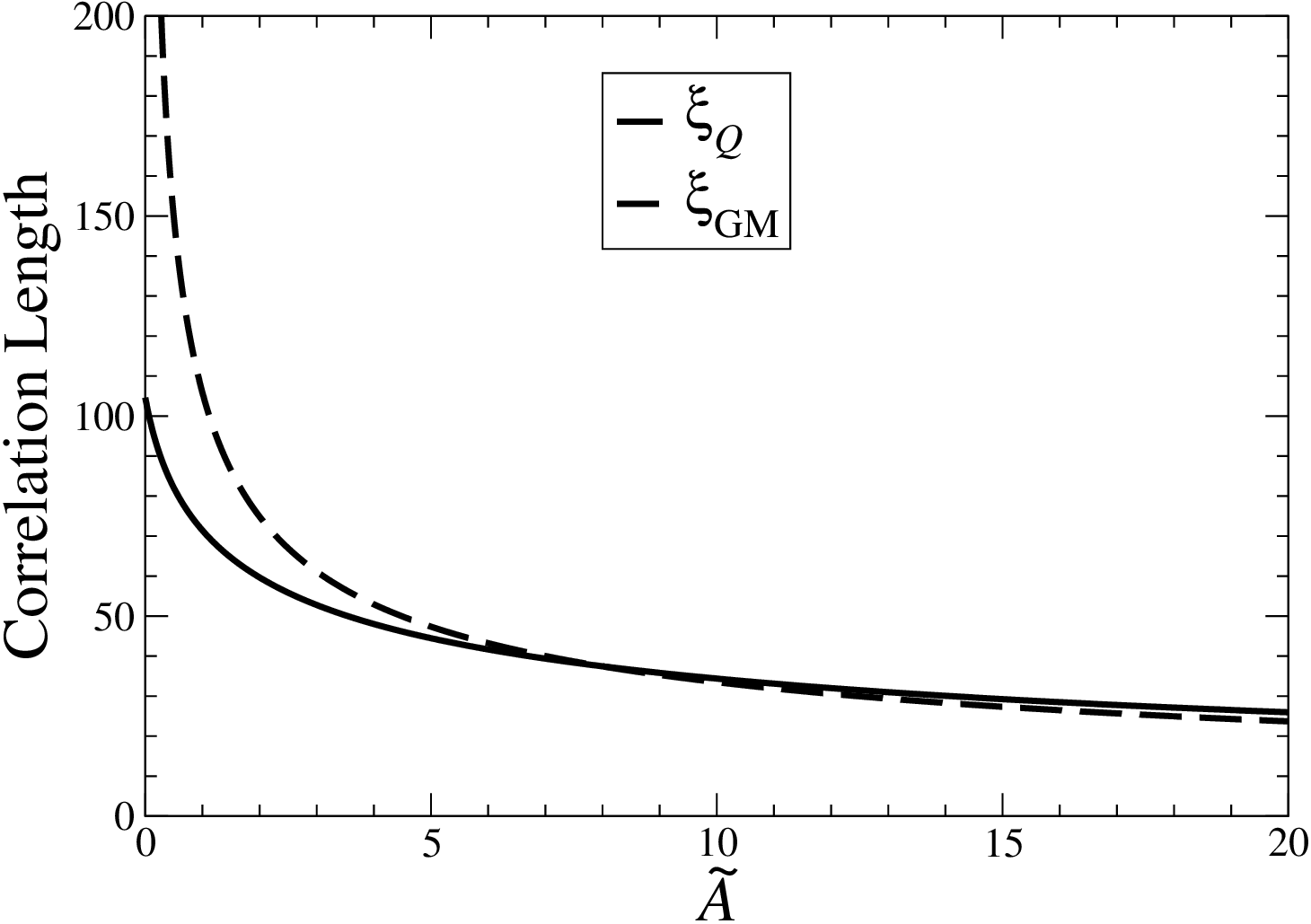}}
\end{center}
\caption[shrt]{
Comparison of the correlation length calculated from the 
SPD ($\xi_Q$) and from the Gruber-Mullins approximation 
($\xi_{\rm GM}$), evaluated numerically for 
$\langle L \rangle \! = \! 10$ and $k_{\rm B}T \! = \! 0.45\epsilon$. 
Although there is decent agreement for 
$\tilde{A} \! > \! 5$, $\xi_{\rm GM}$ unphysically diverges as 
$\tilde{A} \! \rightarrow \!  0$.  In contrast, $\xi_Q$
remains finite and reasonable for all nonegative
values of $\tilde{A}$.
}
\protect\label{fig:xi}
\end{figure}

The observation, made in the previous section, that $Q(x)$ is to a very 
good approximation Gaussian is helpful towards the calculation of 
the characteristic length for $\sigma^2_Q$.  In Ref.~\onlinecite{Bartelt90},
the ``TWD'' was calculated within the Gruber-Mullins approximation; because 
the position of only one step was explicitly taken into account, though, 
it could equally be considered a SPD.  In fact, the Gaussian solution is 
a more appropriate description of a SPD, which is symmetric, than a TWD, 
which is asymmetric. Substituting the variance of the SPD into 
Eq.~(18) of Ref.~\onlinecite{Bartelt90}, we find the correlation length to 
be  
\begin{equation}
  \label{eq:xi}
   \xi_Q = \frac{2\langle L \rangle^2 \tilde{\beta}\sigma^2_{Q,{\rm W}}}
                {k_{\rm B}T} \, .
\end{equation} 
Figure~\ref{fig:xi} shows a comparison between $\xi_Q$ and the correlation 
length from Ref.~\onlinecite{Bartelt90}.  

\begin{figure}
\epsfxsize=5.7cm
\begin{center}
\centerline{\epsfbox{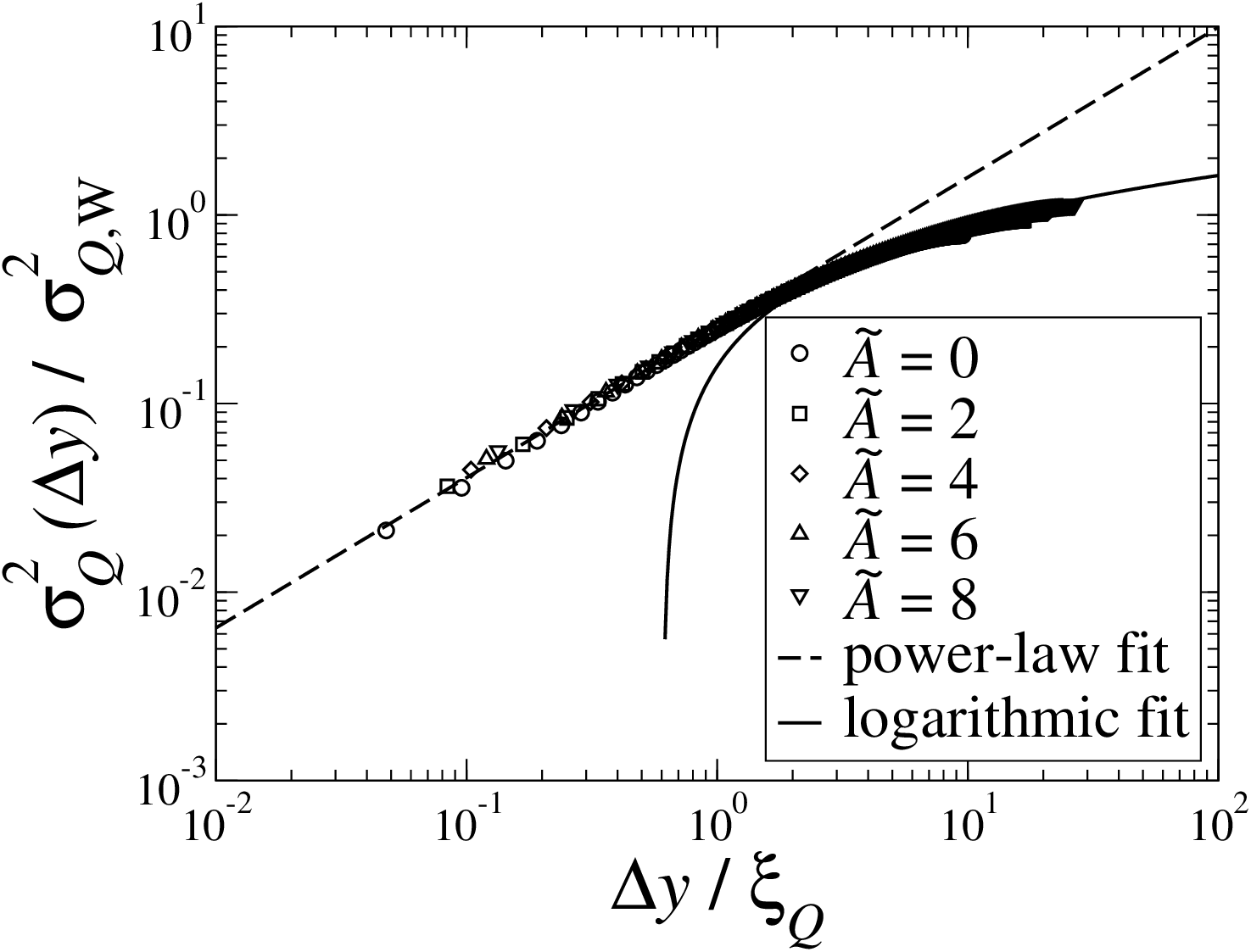}}
\end{center}
\caption[shrt]{
A power-law fit to all the Monte Carlo estimates of 
$\sigma^2_Q(\Delta y)$ for $0 \! \leq \! \tilde{A} \! \leq \! 8$, 
$\Delta y \! < \! \xi_Q$, 
indicates an initial growth with an exponent of 
$0.797 \pm 0.006$.
}
\protect\label{fig:power_spdvar}
\end{figure}

\begin{figure}
\epsfxsize=5.7cm
\begin{center}
\centerline{\epsfbox{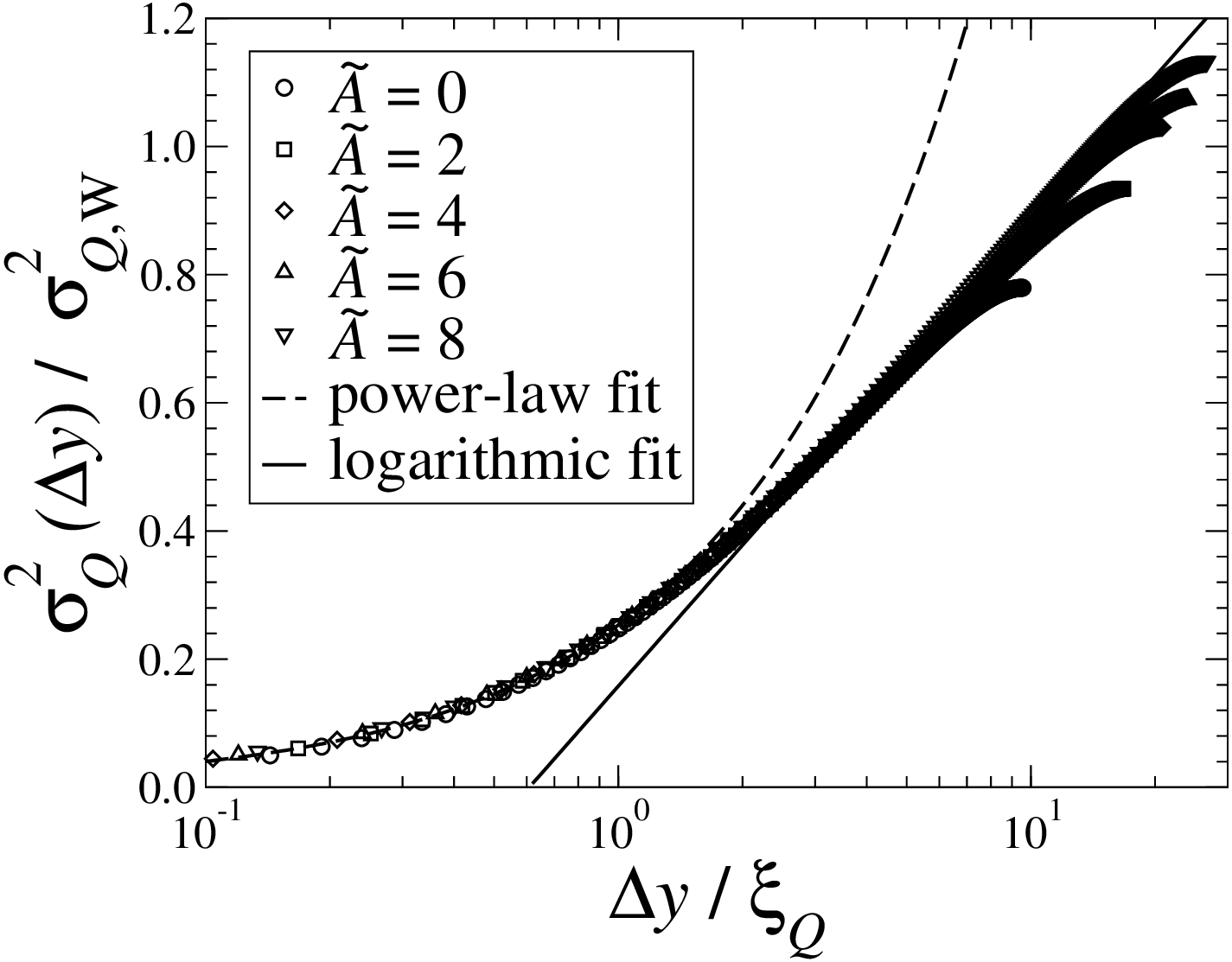}}
\end{center}
\caption[shrt]{
A fit to the Monte Carlo estimates of
$\sigma^2_Q(\Delta y)$ for $\tilde{A} \! = \! 8$, 
$4\xi_Q \! < \! \Delta y \! < \! L_y/2$, 
indicates an asymptotic growth given by 
$\sigma^2_Q(\Delta y)/\sigma^2_{Q,{\rm W}} 
  \! \approx \! 0.158 + 0.3175 \ln (\Delta y/\xi_Q)$.
The length $L_{\rm W}$, defined by Eq.~(\protect\ref{eq:defLW}),
$\sigma^2_Q(L_{\rm W}) \! \equiv \! \sigma^2_{Q,{\rm W}}$, 
is consequently given by 
$L_{\rm W} \! = \! (14.183 \pm 0.035) \xi_Q$. 
}
\protect\label{fig:log_spdvar}
\end{figure}

Scaled by $\sigma^2_{Q,{\rm W}}$ and $\xi_Q$, 
$\sigma^2_Q(\Delta y)$ appears to be independent of $\tilde{A}$; 
although the PEM incorrectly predicts 
that $\sigma^2_Q(\Delta y)$ remains finite in the limit 
$\Delta y \! \rightarrow \! \infty$, it nevertheless provides the 
correct scaling factors.  This is remarkable, since although 
$g_x(\Delta y)$ shows scaling with temperature, it 
does not exhibit scaling independent of $\tilde{A}$\cite{role}.

For $\Delta y \! < \! \xi_Q$, a least-squares fit indicates
power-law growth of $\sigma^2_Q(\Delta y)$ with an exponent of 
$0.797 \pm 0.006$ (Fig.~\ref{fig:power_spdvar}).  
Equation~(\ref{eq:approxsig2}) predicts 
power-law growth, but with an exponent of 1.  Interestingly, 
the power-law behavior of $g_x(\Delta y)$ extends only out to\cite{role} 
$\Delta y \! \approx \! 0.1 y_{\rm coll}$; since 
$y_{\rm coll} \! = \! \xi_Q / (\pi -2)$ 
(for $\tilde{A} \! = \! 0$), power-law scaling extends farther 
for $\sigma^2_Q(\Delta y)$ than for $g_x(\Delta y)$.

For large $\Delta y$, $\sigma^2_Q(\Delta y)$ follows 
the logarithmic scaling of Eq.~(\ref{eq:logscaledsig2}).  
A least-squares fit was performed on the $\tilde{A} \! = \! 8$ data,  
since this has the smallest value of $\xi_Q$ among the available 
simulations, and hence the largest available values of $\Delta y/\xi_Q$.  To 
avoid the crossover from the power-law regime, the fit was restricted to 
$\Delta y \! > \! 4\xi_Q$; likewise, the fit was limited to 
$\Delta y \! < \! L_y/2$ to limit finite-size effects. 
The resulting fit, shown in Fig.~\ref{fig:log_spdvar}, is in good agreement 
with data from all values of $\tilde{A}$ except where finite-size effects 
become evident. The fitted parameters, 
$c_4 \! = \! 0.1578 \pm 0.0004$ and $c_5 \! = \! 0.3175 \pm 0.0002$, 
allow us to find a ``Wigner length'', $L_{\rm W}$, defined by 
\begin{equation}
  \label{eq:defLW}
   \sigma^2_Q(L_{\rm W}) \equiv \sigma^2_{Q,{\rm W}} \, , 
\end{equation}
{}to be 
\begin{equation}
  \label{eq:LWval}
   L_{\rm W}  =  (14.183 \pm 0.035) \xi_Q \, .  
\end{equation}

\section{Scaling of the TWD}
\label{sec-TWDs}

It seems somewhat surprising that so many correlation 
lengths are necessary for the PEM to agree with the observed 
variance.  To better understand this,  
it is helpful to consider the corresponding scaling 
of the TWD when it is calculated under the same restrictions as 
$\sigma^2_Q(\Delta y)$.  Specifically, the TWD must be averaged 
over a given length, $\Delta y$, of a {\em single pair of adjacent 
steps in a single ``snapshot''}.  This is very different from the 
analysis presented in Ref.~\onlinecite{Hailu04}, where, as in 
other previous work, averages were made over the entire length $L_y$ 
of the simulations, over all pairs of neighboring steps, and over 
all ``snapshots''.  Remembering that the $y$-direction 
corresponds to time in the worldline interpretation of steps, 
the averages we are about to calculate correspond to time averages 
in statistical mechanics, whereas the previous averages have combined 
the time average with two kinds of ensemble average (over different 
pairs and different ``snapshots'').  Only in the limits of long 
times and large ensembles should one expect these averages to be 
identical\cite{Pathria}. 

\begin{figure}
\epsfxsize=5.7cm
\begin{center}
\centerline{\epsfbox{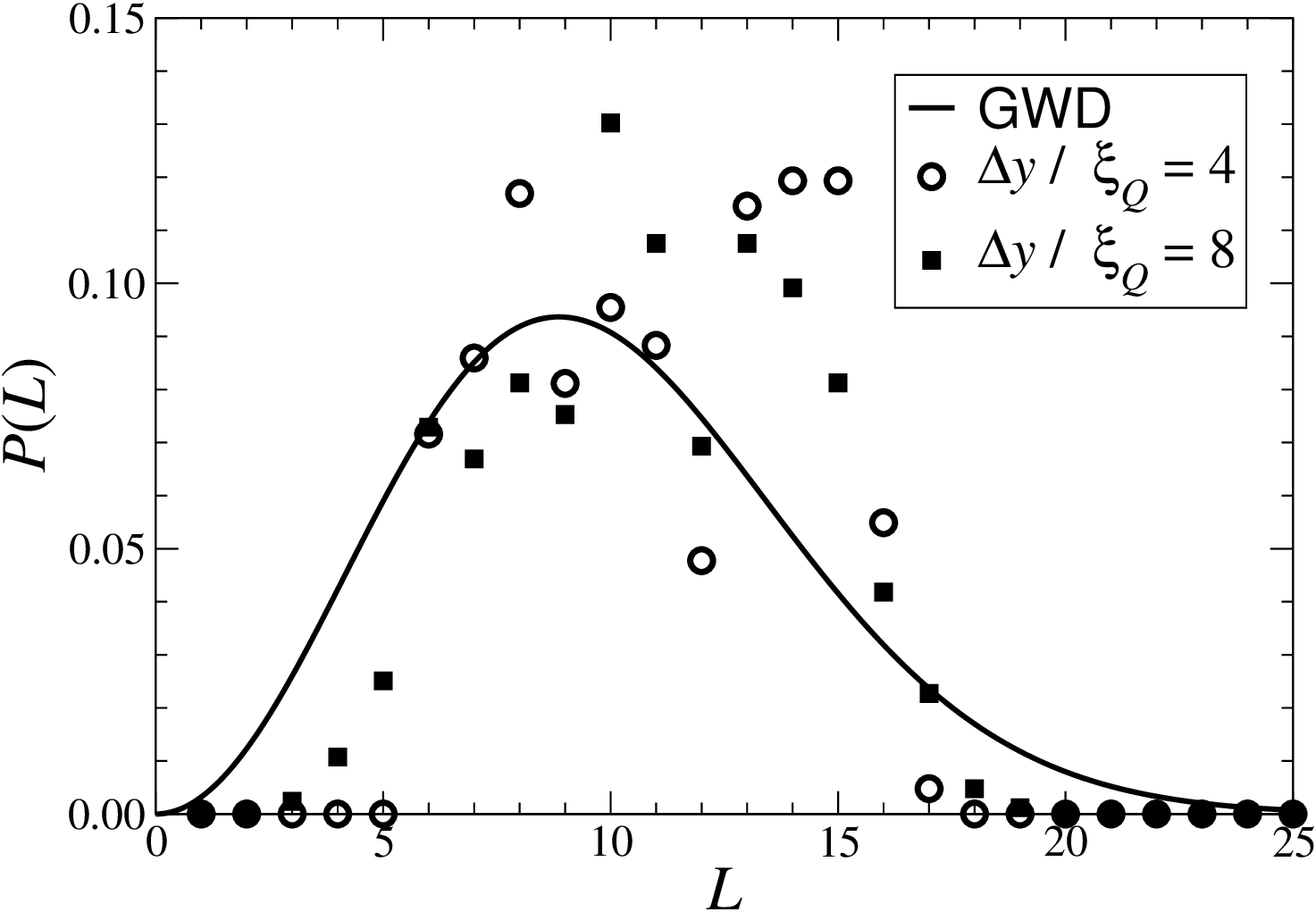}}
\end{center}
\caption[shrt]{
Terrace Width Distributions calculated between a single pair 
of neighboring steps depend on $\Delta y$, the length of 
step over which the distribution is averaged.  Although in the 
limit $\Delta y \rightarrow \infty$ the TWD converges to 
the Generalized Wigner Distribution (to a very good approximation), 
when $\Delta y / \xi_Q$ is small the TWD is dominated by noise.  
These results are typical for $\tilde{A} \! = \! 0$. 
}
\protect\label{fig:Atilde00TWD}
\end{figure}

\begin{figure}
\epsfxsize=5.7cm
\begin{center}
\centerline{\epsfbox{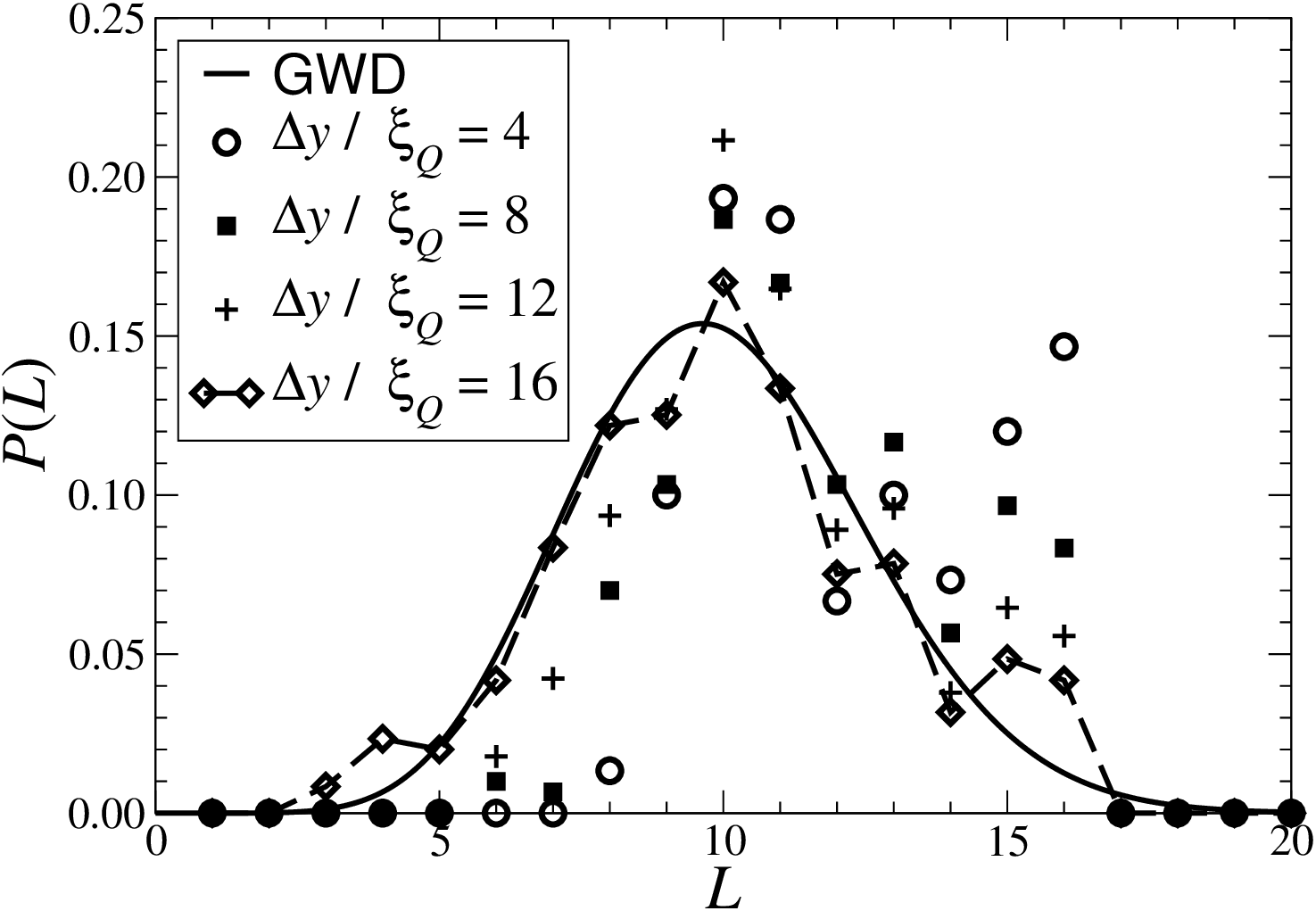}}
\end{center}
\caption[shrt]{
For small $\Delta y / \xi_Q$, typical TWDs are dominated by 
noise, but for 
$\Delta y \! > \! L_{\rm W} \! \approx \! 14.2 \xi_Q$, 
the Generalized Wigner Distribution dominates. 
These results are for $\tilde{A} \! = \! 8$. 
}
\protect\label{fig:Atilde08TWD}
\end{figure}

\begin{figure}
\epsfxsize=5.7cm
\begin{center}
\centerline{\epsfbox{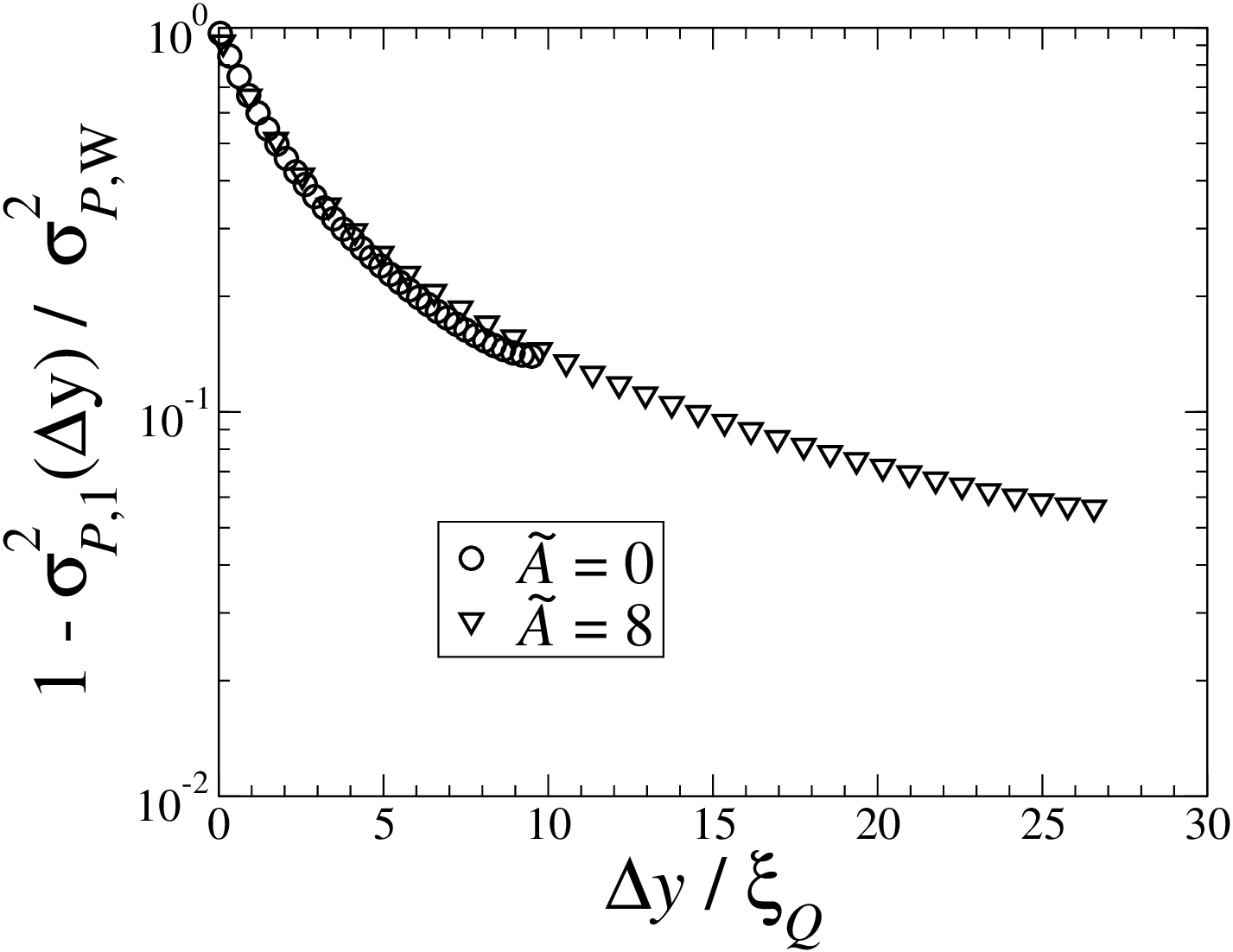}}
\end{center}
\caption[shrt]{
The variance of the TWD also approximately scales with
$\Delta y / \xi_Q$.    At  
$\Delta y \! = \! L_{\rm W} \! \approx \! 14.2 \xi_Q$, 
the average variance of TWDs generated from single pairs 
of neighboring steps is within about \%10 
of the variance given by the GWD. 
}
\protect\label{fig:scaled_twdvar}
\end{figure}

In the language of Ref.~\onlinecite{role}, $\xi_Q$ is 
approximately the distance between ``collisions'' of 
neighboring steps.  In order to sample the distribution of 
terrace widths adequately, a step must ``collide'' several 
times with its neighbors.  This is shown clearly in 
Figs.~\ref{fig:Atilde00TWD} and \ref{fig:Atilde08TWD}. 
For the simulation parameters given in Sec.~\ref{sec-numerical}, 
$L_{\rm W} \! > \! L_y$ for $\tilde{A} \! = \! 0$; consequently, 
the TWDs shown in Fig.~\ref{fig:Atilde00TWD} are dominated by 
noise.  For $\tilde{A} \! = \! 8$, on the other hand, 
$L_{\rm W} \! < \! L_y$, and we are able to see 
in Fig.~\ref{fig:Atilde08TWD} the crossover into 
the regime where the GWD, not noise, is dominant. 

The role of step collisions in equilibrating the TWD can also 
be seen from $\sigma^2_P(\Delta y)$, which is the variance of TWDs
calculated from a length $\Delta y$ of neighboring steps, 
averaged over all pairs of neighboring steps, starting points, 
and ``snapshots''.  Like $\sigma^2_Q(\Delta y)$, 
$\lim_{\Delta y \rightarrow 0} \sigma^2_P(\Delta y) \! = \! 0$; 
unlike $\sigma^2_Q(\Delta y)$, 
$\lim_{\Delta y \rightarrow \infty} \sigma^2_P(\Delta y)$ 
is finite and given approximately\cite{Hailu04} by 
the PEM result\cite{EP99},  
\begin{equation}
  \label{eq:sigma2TWD} 
   \sigma^2_{P,{\rm W}} = \frac{\varrho + 1}{2 b_\varrho} - 1 \, . 
\end{equation}
This suggests plotting 
$1 - \sigma^2_P(\Delta y)/\sigma^2_{P,{\rm W}}$ vs.\ 
$\Delta y/\xi_Q$ to determine whether the approach to the asymptotic 
limit is exponential or power-law.  As shown in 
Fig.~\ref{fig:scaled_twdvar}, the scaling appear to be 
neither a simple power law nor a simple exponential decay, 
but it is difficult to be certain since the TWD does not 
converge {\em exactly} to the GWD 
even in the limit $\Delta y \rightarrow \infty$.
Also, the scaling does not appear to be quite as precise as in 
Figs.~\ref{fig:power_spdvar} and \ref{fig:log_spdvar}.
this is not surprising, since the correlation length 
$\xi_P$ for the TWD is not identical to $\xi_Q$.  

More significantly, Fig.~\ref{fig:scaled_twdvar} indicates that 
$\sigma^2_P(L_{\rm W})$ is within about \%10 of the approximate 
asymptotic value, $\sigma^2_{P,{\rm W}}$.  
This is a very plausible threshold for statistics from
the TWD.

\section{Conclusion}
\label{sec-conclusions}

For the common case in which steps on a vicinal crystal surface interact 
according to Eq.~(\ref{eq:InvSquareInteraction}), the 
generalized Wigner distribution (GWD) has been 
shown previously to be in excellent agreement with the 
terrace width distribution (TWD). To fully 
appreciate the model which predicts the GWD, though, it is 
necessary to examine its predictions for other statistical properties 
and how well these predictions agree with actual measurements.  This 
article has made such a comparison between the predicted and measured 
step position distribution (SPD).  The results demonstrate both the 
strength and limitations of the Pairwise Einstein Model (PEM). 

Since the SPD is so well approximated by a Gaussian, it is tempting to
compare it directly with Gaussian theories of the TWD.  As can be seen in
Table~1, in the limit of strongly interacting steps the variance of the
SPD is slightly larger than that of the Gruber-Mullins approximation,
but less than the variance of the TWD given by either the ``Saclay'' or
``modified Grenoble'' approximations.  This is reasonable; unlike the
Gruber-Mullins Hamiltonian, Eq.~(\ref{eq:HamX1X2}) does not have fixed walls,
so the steps can experience larger fluctuations.  In spite of this, since
the Gruber-Mullins approximation allows only one step to move, it can be
regarded equally as an approximation for the TWD or for the SPD.
The fact that the SPD is smaller than the other approximations of the
TWD is apparently due to the fact that correlations between fluctuations
of adjacent steps are to some degree taken into account in all these
approximations, so that they are specifically approximations for the TWD,
not the SPD.

\begin{center}
\begin{table}
 \caption[Asym]{Asymptotic variances in the limit of strong step-step repulsion.
          The Gaussian-like approximation for the step position distribution
          (SPD) given by Eq.~(\protect\ref{eq:Q}) is compared with selected
          approximations for the TWD. Except for the Generalized Wigner
      Distribution, all approximate TWDs are Gaussian approximations.
          Note also that our approximation for the SPD and the Generalized
          Wigner Distribution are both independent of the number of
          interacting steps, whereas the Gaussian approximations are not.
          (See also Table~1 of Ref.~\protect\cite{SurfSci493_460}.)
 }
 \label{tabl:variances}
 \begin{tabular}[b]{|lcc|} \hline
 Distribution  &  Reference     & Asymptotic Variance \\ \hline
 SPD           & Eq.~(\protect\ref{eq:sigma2}) &  $0.375 \varrho^{-1}$
 \\
 Generalized Wigner 
               & \protect\cite{EP99,MehtaRanMat,Haake}
                                               & $0.5 \varrho^{-1}$ \\
 Gruber Mullins (all steps)
               & \protect\cite{Gruber67}
                                               & $0.278 \varrho^{-1}$ \\
 {\tt "} (nearest neighbors)
               & {\tt "}
                                               & $0.289 \varrho^{-1}$ \\
 Modified Grenoble (all steps)
               & \protect\cite{EP99,PM98,IMP98}
                                               & $0.495 \varrho^{-1}$ \\
 {\tt "} (nearest neighbors)
               & {\tt "} 
                                               & $0.520 \varrho^{-1}$ \\
 Saclay (all steps)
               & \protect\cite{Masson94,Barbier96,LeGoff99}
                                               & $0.405 \varrho^{-1}$
\\ \hline
 \end{tabular}
 \end{table}
\end{center}

Because the PEM confines both steps within a harmonic well, 
the theoretical asymptotic variance of the SPD must be finite.
However, the vicinal surface is rough, and the variance of the 
SPD diverges logarithmically with the length of step $\Delta y$ from which 
it is calculated.  At some finite length, $L_{\rm W}$, the 
prediction of the PEM is accurate.  As Fig.~\ref{fig:log_spdvar} 
shows, $L_{\rm W} \! \approx \! 14.2 \xi_Q$.  That so many 
``collisions'' between neighboring steps are needed to adequately 
sample the statistics resulting from their interactions is supported 
by obeservations of the dependence of the TWD on $\Delta y$, as shown in 
Figs.~\ref{fig:Atilde00TWD}, \ref{fig:Atilde08TWD}, and 
\ref{fig:scaled_twdvar}.  

In principle, the SPD could be used to determine $\tilde{A}$.  However, 
the SPD is strongly affected by the random position of the
average step position, and it depends far too strongly on $\Delta y$. 
The TWD has neither of these two restrictions and is a more practical 
alternative for determining $\tilde{A}$.
Instead, the utility of the SPD lies in clarifying the Pairwise Einstein Model. 
The finite length $L_{\rm W}$ introduced in this work emerges more naturally 
than the finite length $\cal V$ that was introduced in Refs.~\onlinecite{beyond}
and \onlinecite{Yancey05}, but the two are obviously related.  Both help 
describe a short-lived dynamic constraint that is roughly analagous to 
a reptation tube\cite{Doi} in polymer physics. 

Naturally, the remarkable success of the Pairwise Einstein Model suggests that 
a Debye model\cite{AshcroftMermin} might lead to even better descriptions of 
vincinal crystal surfaces.  Preliminary results\cite{Greene06} from such studies 
correctly show that $g_x(\Delta y)$ diverges logarithmically. 

\section*{Acknowledgment}

This research was supported by an award from Research Corporation.
The authors also thank Jeremy Yancey and April St.~John for
critical readings.

\bibliographystyle{prsty}


\end{document}